\documentclass[aps,prl,twocolumn,showpacs,groupedaddress]{revtex4}  % for review and submission 
\usepackage{graphicx}  % needed for figures
\usepackage{dcolumn}   % needed for some tables
\usepackage{bm}        % for math
\usepackage{amssymb}   % for math

\begin{document}
% the following information is for internal review, please remove them for submission
%\leftline{Version 0.8 as of \today} 
%\leftline{Primary authors: C. Gerber, T. Golling, G. Otero, M.-A. Pleier,
%E. Shabalina and J-R. Vlimant}
%\rightline{Comment to {\tt d0-run2eb-009@fnal.gov} by April 11, 2005}
\hspace{5.2in} \mbox{FERMILAB-PUB-05-079-E} 
%\leftline{}
%\leftline{}
%\leftline{}
%\rightline {\mbox{Fermilab-Pub-xx/xxx-E}}
%\rightline {}
\newcommand{\dzero}     {D\O}
\newcommand{\ttbar}     {\mbox{$t\bar{t}$}}
\newcommand{\met}       {\mbox{$\not\!\!E_T$}}
\newcommand{\pythia}    {\sc{pythia}}
\newcommand{\alpgen}    {\sc{alpgen}}
\newcommand{\geant}    {\sc{geant}}
\newcommand{\ljets}    {\mbox{$\ell$+jets}}
\newcommand{\ejets}    {\mbox{$e$+jets}}
\newcommand{\mujets}    {\mbox{$\mu$+jets}}
 
\title{ Measurement of the $t\overline{t}~$ Production Cross Section in
{\mbox{$p\bar p$}}\
Collisions at {\mbox{$\sqrt{s}$ =\ 1.96\ TeV}} using Kinematic 
Characteristics of Lepton + Jets Events}

% LIST_OF_AUTHORS_R2.TEX                 4/8/05             
%
\author{                                                                      
%% names begin here                                                           
V.M.~Abazov,$^{35}$                                                           
B.~Abbott,$^{72}$                                                             
M.~Abolins,$^{63}$                                                            
B.S.~Acharya,$^{29}$                                                          
M.~Adams,$^{50}$                                                              
T.~Adams,$^{48}$                                                              
M.~Agelou,$^{18}$                                                             
J.-L.~Agram,$^{19}$                                                           
S.H.~Ahn,$^{31}$                                                              
M.~Ahsan,$^{57}$
% ...
\\                                                                            
\vskip 0.30cm                                                                 
\centerline{(D\O\ Collaboration)}                                             
\vskip 0.30cm                                                                 
}                                                                             
\affiliation{                                                                 
\centerline{$^{1}$Universidad de Buenos Aires, Buenos Aires, Argentina}       
\centerline{$^{2}$LAFEX, Centro Brasileiro de Pesquisas F{\'\i}sicas,         
                  Rio de Janeiro, Brazil}                                     
\centerline{$^{3}$Universidade do Estado do Rio de Janeiro,                   
                  Rio de Janeiro, Brazil}                                     
\centerline{$^{4}$Instituto de F\'{\i}sica Te\'orica, Universidade            
                  Estadual Paulista, S\~ao Paulo, Brazil}                     
\centerline{$^{5}$University of Alberta, Edmonton, Alberta, Canada,           
               Simon Fraser University, Burnaby, British Columbia, Canada,}   
\centerline{York University, Toronto, Ontario, Canada, and                    
         McGill University, Montreal, Quebec, Canada}                         
\centerline{$^{6}$Institute of High Energy Physics, Beijing,                  
                  People's Republic of China}                                 
\centerline{$^{7}$University of Science and Technology of China, Hefei,       
                  People's Republic of China}                                 
\centerline{$^{8}$Universidad de los Andes, Bogot\'{a}, Colombia}             
\centerline{$^{9}$Center for Particle Physics, Charles University,            
                  Prague, Czech Republic}                                     
\centerline{$^{10}$Czech Technical University, Prague, Czech Republic}        
\centerline{$^{11}$Institute of Physics, Academy of Sciences, Center          
                  for Particle Physics, Prague, Czech Republic}               
\centerline{$^{12}$Universidad San Francisco de Quito, Quito, Ecuador}        
\centerline{$^{13}$Laboratoire de Physique Corpusculaire, IN2P3-CNRS,         
                 Universit\'e Blaise Pascal, Clermont-Ferrand, France}        
\centerline{$^{14}$Laboratoire de Physique Subatomique et de Cosmologie,      
                  IN2P3-CNRS, Universite de Grenoble 1, Grenoble, France}     
\centerline{$^{15}$CPPM, IN2P3-CNRS, Universit\'e de la M\'editerran\'ee,     
                  Marseille, France}                                          
\centerline{$^{16}$Laboratoire de l'Acc\'el\'erateur Lin\'eaire,              
                  IN2P3-CNRS, Orsay, France}                                  
\centerline{$^{17}$LPNHE, IN2P3-CNRS, Universit\'es Paris VI and VII,         
                  Paris, France}                                              
\centerline{$^{18}$DAPNIA/Service de Physique des Particules, CEA, Saclay,    
                  France}                                                     
\centerline{$^{19}$IReS, IN2P3-CNRS, Universit\'e Louis Pasteur, Strasbourg,  
                France, and Universit\'e de Haute Alsace, Mulhouse, France}   
\centerline{$^{20}$Institut de Physique Nucl\'eaire de Lyon, IN2P3-CNRS,      
                   Universit\'e Claude Bernard, Villeurbanne, France}         
\centerline{$^{21}$III. Physikalisches Institut A, RWTH Aachen,               
                   Aachen, Germany}                                           
\centerline{$^{22}$Physikalisches Institut, Universit{\"a}t Bonn,             
                  Bonn, Germany}                                              
\centerline{$^{23}$Physikalisches Institut, Universit{\"a}t Freiburg,         
                  Freiburg, Germany}                                          
\centerline{$^{24}$Institut f{\"u}r Physik, Universit{\"a}t Mainz,            
                  Mainz, Germany}                                             
\centerline{$^{25}$Ludwig-Maximilians-Universit{\"a}t M{\"u}nchen,            
                   M{\"u}nchen, Germany}                                      
\centerline{$^{26}$Fachbereich Physik, University of Wuppertal,               
                   Wuppertal, Germany}                                        
\centerline{$^{27}$Panjab University, Chandigarh, India}                      
\centerline{$^{28}$Delhi University, Delhi, India}                            
\centerline{$^{29}$Tata Institute of Fundamental Research, Mumbai, India}     
\centerline{$^{30}$University College Dublin, Dublin, Ireland}                
\centerline{$^{31}$Korea Detector Laboratory, Korea University,               
                   Seoul, Korea}                                              
\centerline{$^{32}$CINVESTAV, Mexico City, Mexico}                            
\centerline{$^{33}$FOM-Institute NIKHEF and University of                     
                  Amsterdam/NIKHEF, Amsterdam, The Netherlands}               
\centerline{$^{34}$Radboud University Nijmegen/NIKHEF, Nijmegen, The          
                  Netherlands}                                                
\centerline{$^{35}$Joint Institute for Nuclear Research, Dubna, Russia}       
\centerline{$^{36}$Institute for Theoretical and Experimental Physics,        
                  Moscow, Russia}                                             
\centerline{$^{37}$Moscow State University, Moscow, Russia}                   
\centerline{$^{38}$Institute for High Energy Physics, Protvino, Russia}       
\centerline{$^{39}$Petersburg Nuclear Physics Institute,                      
                   St. Petersburg, Russia}                                    
\centerline{$^{40}$Lund University, Lund, Sweden, Royal Institute of          
                   Technology and Stockholm University, Stockholm,            
                   Sweden, and}                                               
\centerline{Uppsala University, Uppsala, Sweden}                              
\centerline{$^{41}$Lancaster University, Lancaster, United Kingdom}           
\centerline{$^{42}$Imperial College, London, United Kingdom}                  
\centerline{$^{43}$University of Manchester, Manchester, United Kingdom}      
\centerline{$^{44}$University of Arizona, Tucson, Arizona 85721, USA}         
\centerline{$^{45}$Lawrence Berkeley National Laboratory and University of    
                  California, Berkeley, California 94720, USA}                
\centerline{$^{46}$California State University, Fresno, California 93740, USA}
\centerline{$^{47}$University of California, Riverside, California 92521, USA}
\centerline{$^{48}$Florida State University, Tallahassee, Florida 32306, USA} 
\centerline{$^{49}$Fermi National Accelerator Laboratory, Batavia,            
                   Illinois 60510, USA}                                       
\centerline{$^{50}$University of Illinois at Chicago, Chicago,                
                   Illinois 60607, USA}                                       
\centerline{$^{51}$Northern Illinois University, DeKalb, Illinois 60115, USA} 
\centerline{$^{52}$Northwestern University, Evanston, Illinois 60208, USA}    
\centerline{$^{53}$Indiana University, Bloomington, Indiana 47405, USA}       
\centerline{$^{54}$University of Notre Dame, Notre Dame, Indiana 46556, USA}  
\centerline{$^{55}$Iowa State University, Ames, Iowa 50011, USA}              
\centerline{$^{56}$University of Kansas, Lawrence, Kansas 66045, USA}         
\centerline{$^{57}$Kansas State University, Manhattan, Kansas 66506, USA}     
\centerline{$^{58}$Louisiana Tech University, Ruston, Louisiana 71272, USA}   
\centerline{$^{59}$University of Maryland, College Park, Maryland 20742, USA} 
\centerline{$^{60}$Boston University, Boston, Massachusetts 02215, USA}       
\centerline{$^{61}$Northeastern University, Boston, Massachusetts 02115, USA} 
\centerline{$^{62}$University of Michigan, Ann Arbor, Michigan 48109, USA}    
\centerline{$^{63}$Michigan State University, East Lansing, Michigan 48824,   
                   USA}                                                       
\centerline{$^{64}$University of Mississippi, University, Mississippi 38677,  
                   USA}                                                       
\centerline{$^{65}$University of Nebraska, Lincoln, Nebraska 68588, USA}      
\centerline{$^{66}$Princeton University, Princeton, New Jersey 08544, USA}    
\centerline{$^{67}$Columbia University, New York, New York 10027, USA}        
\centerline{$^{68}$University of Rochester, Rochester, New York 14627, USA}   
\centerline{$^{69}$State University of New York, Stony Brook,                 
                   New York 11794, USA}                                       
\centerline{$^{70}$Brookhaven National Laboratory, Upton, New York 11973, USA}
\centerline{$^{71}$Langston University, Langston, Oklahoma 73050, USA}        
\centerline{$^{72}$University of Oklahoma, Norman, Oklahoma 73019, USA}       
\centerline{$^{73}$Brown University, Providence, Rhode Island 02912, USA}     
\centerline{$^{74}$University of Texas, Arlington, Texas 76019, USA}          
\centerline{$^{75}$Southern Methodist University, Dallas, Texas 75275, USA}   
\centerline{$^{76}$Rice University, Houston, Texas 77005, USA}                
\centerline{$^{77}$University of Virginia, Charlottesville, Virginia 22901,   
                   USA}                                                       
\centerline{$^{78}$University of Washington, Seattle, Washington 98195, USA}  
}                                                                             
%end                                                                          

\date{April 25, 2005}

\begin{abstract}
We present a measurement of the top quark pair ($\ttbar$)  
production cross section 
($\sigma_{t\overline{t}}$) 
in {\mbox{$p\bar p$}} collisions at a center-of-mass energy of 1.96 TeV 
using $230\;\rm pb^{-1}$ of data collected by the D\O\ detector at
the Fermilab Tevatron Collider. 
We select events with one charged lepton 
(electron or muon), large missing transverse energy, and at least four jets,
and extract the $t\bar{t}$ content of the sample based on the kinematic  
characteristics of the events. 
For a top quark mass of 175 GeV, we measure
$
\sigma_{t\overline{t}} = 
6.7^{+1.4}_{-1.3}\:{\rm (stat)}\: 
^{+ 1.6}_{- 1.1}\:{\rm (syst)}\:                
\pm0.4\:{\rm(lumi)}\:{\rm pb},$
\noindent in good agreement with the standard model prediction.
%\vspace{1cm}
\end{abstract}

\pacs{13.85.Lg, 13.85.Qk, 14.65.Ha}
\maketitle
%\vspace{5cm}

Within the standard model (SM),
top quarks are produced in $p\bar{p}$ collisions predominantly in pairs 
via the strong interaction ($q\bar{q}$ annihilation and
gluon fusion), and decay almost exclusively to a $W$ boson and a $b$ quark. 
The top quark pair production cross section 
$\sigma_{t\overline{t}}$ was measured by the 
CDF~\cite{runIcdf} 
and \dzero~\cite{runI} collaborations at a center-of-mass energy of 1.8 TeV. 
Recent measurements~\cite{cdf} of $\sigma_{t\overline{t}}$
at {\mbox{$\sqrt{s}$ =\ 1.96\ TeV}} have focused
on the selection of candidates via the reconstruction of
displaced vertices signaling the presence of $b$ quarks in the
final state. These measurements assume that the branching ratio of the
top quark $B (t \to Wb) = 1$, thus making an
implicit use of the SM prediction that $|V_{tb}|=0.9990 \div 0.9992$ (at 90\%
C.L.)~\cite{PDG2002}. This
prediction is based on the requirements that there are three fermion families 
and the CKM matrix is unitary. If these assumptions are relaxed,
$|V_{tb}|$ is essentially unconstrained, which
allows for large deviations of $B (t \to Wb)$ from unity \cite{branch}. Such
deviations would be an indication of physics beyond the SM.
Our analysis exploits only the kinematic properties of the events to 
separate signal from background, with no assumptions about the 
multiplicity of final-state $b$ quarks,
thus providing a less model-dependent determination
of the top quark production cross section.

In this Letter, we report a new measurement of  $\sigma_{t\overline{t}}$
using data collected with the D\O\ detector
from August 2002 through March 2004 at the Fermilab Tevatron
{\mbox{$p\bar p$}}\ collider at {\mbox{$\sqrt{s}$ =\ 1.96\ TeV}}. 
The decay channel used in this analysis is 
{\mbox{$ t\overline{t} \rightarrow W^{+} W^{-} q\overline{q}$}}, with the
subsequent decay of one $W$ boson into two quarks, and the other  
$W$ boson into a charged lepton and
a neutrino. We refer to this decay mode of $t\overline{t}$ events as the
lepton+jets ($\ell$+jets) channel. These events 
are characterized by the presence of one high-$p_T$ isolated 
electron ($\ejets$ channel) or muon ($\mujets$ channel), large transverse 
energy imbalance due to the undetected 
neutrino ({\mbox{$\not\!\!E_T$}}), and at least four hadronic jets.  

The three main subsystems of the D\O\ Run II detector~\cite{d0det} 
used in this analysis are the
central tracking system, the liquid-argon/uranium calorimeters, and the muon
spectrometer. The central tracking system is located within a 
2 Tesla superconducting solenoidal magnet, and consists of a silicon
microstrip tracker (SMT) and a central fiber tracker (CFT) that provide
tracking and vertexing in the pseudorapidity~\cite{eta} range 
$|\eta|<3.0$. The primary interaction vertex of the events was 
required to be within $60\;\rm cm$ of the center of the detector 
along the direction of the beam.
Electrons and jets were detected in hermetic calorimeters~\cite{EMNIM,TBNIM} 
with transverse granularity $\Delta\eta \times 
\Delta\phi = 0.1 \times 0.1$, where 
$\phi$ is the azimuthal angle. The third layer of the 
electromagnetic (EM) calorimeter, in which the maximum energy 
deposition of EM showers is expected, has a finer  
granularity $\Delta\eta \times \Delta\phi = 0.05 \times 0.05$.
The calorimeters consist of a central 
section (CC) covering the region  $|\eta|<1.1$, and two end 
calorimeters (EC) extending coverage to $|\eta| \approx 4.2$.  
Muons were detected as tracks reconstructed from hits recorded  
in three layers of tracking detectors and two layers of 
scintillators~\cite{MUONNIM}, both located outside the calorimeter. 
A $1.8\;\rm Tesla$ iron toroidal magnet is located outside the innermost layer
of the muon detector. 
%The muon momentum resolution in this analysis was
%$\sigma(p_T) = 0.02~\oplus~0.002 p_T$ (with $p_T$ in GeV).
The luminosity was calculated by measuring the rate 
for {\mbox{$p\bar p$}}\ inelastic collisions using two 
hodoscopes of scintillation counters mounted close to the
beam pipe on the front surfaces of the EC calorimeters.
 
Jets were defined using a cone algorithm \cite{jet_algo} with radius 
$\Delta {\cal{R}} = \sqrt {(\Delta\eta)^2+(\Delta\phi)^2} = 0.5$. To 
improve calorimeter performance we use an algorithm that supresses cells with
negative energy as well as cells with energies significantly below the
average electronics noise (unless they neighbor a cell of high 
positive energy).     
%The offline jet identification requirements
%consisted of the following: {\it i}) the jet had to 
%deposit between 5\% and 95\% of its energy in the electromagnetic 
%calorimeter and less than 40\% of its energy in the outermost hadronic section 
%of the calorimeter; {\it ii}) the ratio of the highest to the next-to-highest 
%transverse energy calorimeter cell in the jet had to be less than 10; 
%{\it iii}) each single calorimeter tower had to contain less than 90\%
%of the jet energy;
%{\it iv}) the jet was required to be confirmed by the independent
%trigger readout.
Identified jets were required to be confirmed by the independent
trigger readout.

In the $\ejets$ channel, 
we accepted electrons with $|\eta|<1.1$ and jets
with rapidity $|y|< 2.5$~\cite{eta}.  
At the trigger level, we required a single electron with transverse
momentum ($p_T$) greater than $15\;\rm GeV$, and a jet with 
$p_T>15\;\rm GeV$ ($20\;\rm GeV$) for the first (second)
half of the data. The total integrated luminosity for this sample is 
$226 \pm 15$ pb$^{-1}$.
The offline electron identification requirements 
consisted of the following:
{\it i}) the electron had to deposit at least 90\% of its energy 
in the electromagnetic calorimeter within a cone of 
radius $\Delta {\cal{R}} =  0.2$ relative to 
the shower axis;
{\it ii}) the electron had to be isolated, i.e., the ratio of the energy in 
the hollow cone $0.2 < \Delta {\cal{R}} < 0.4$ to the reconstructed 
electron energy could not exceed 15\%;
{\it iii}) the transverse and longitudinal shower shapes had to be 
consistent with those expected for an electron (based on a detailed Monte 
Carlo simulation); and 
{\it iv}) a good spatial match had to exist between a reconstructed track in 
the tracking system and the shower position in the calorimeter.  
Electrons satisfying the above requirements are referred to as ``loose.''
For a ``tight'' electron, we required in addition, that a
discriminant formed by combining the above variables with the information about 
impact parameter of the matched track relative to the primary interaction
vertex, and the number and $p_T$ of other tracks around the electron candidate,    
be consistent with the expectations for a high-$p_T$ isolated electron.

In the $\mu+$jets channel, we accepted muons with \mbox{${|\eta|<2.0}$} 
and jets
with $|y|<2.5$.
At the trigger level, we required
a single muon detected outside the toroidal magnet 
(which corresponds to an effective minimum momentum of $\approx 3\;\rm GeV$), 
and a jet with
$p_T>20\;\rm GeV$ ($25\;\rm GeV$) for the first (second)
half of the data.
The total integrated luminosity for this sample is $229 \pm 15$ pb$^{-1}$.
The offline muon identification requirements consisted of the 
following:
{\it i}) 
a muon track segment on the inside of the toroid had to be matched to a muon 
track segment on the outside of the toroid; 
{\it ii}) the timing of the muon, based on information from associated 
scintillator 
hits, had to be inconsistent with that of a cosmic ray; 
{\it iii}) a track reconstructed in the tracking system and
pointing to the event vertex was required to be matched to the muon candidate 
found in the muon system; 
{\it iv}) the reconstructed muon was required to be separated from jets, 
$\Delta {\cal{R}}(\mu,{\rm jet})>0.5$.   
Muons satisfying the above requirements are referred to as ``loose.''
For a ``tight'' muon we also applied  
a stricter isolation requirement based on the 
energy of calorimeter clusters and tracks around the muon candidate. 

We selected 87 (80) events that had only one tight electron (muon) 
with $p_T > 20\;\rm GeV$, $\met> 20\;\rm GeV$ and not collinear with the lepton 
direction in the transverse plane, and  
at least four jets each with $p_T > 20$ GeV. We refer to these as the
``tight'' samples in the $e+$jets ($\mu+$jets) channel. 
Removing the tight requirement on the lepton identification results in 
230 (140) events passing the selection. We refer to these as the
``loose'' samples in the $e+$jets ($\mu+$jets) channel. 

Monte Carlo simulations of $\ttbar$ and $W+$jets events were 
used to calculate selection efficiencies and to simulate
kinematic characteristics of the events. 
Top quark signal and $W$+jets background processes were generated at
$\sqrt{s}=1.96$~TeV using {\alpgen}~{\sc 1.2}~\cite{alpgen} for the
parton-level process, and {\pythia}~{\sc 6.2}~\cite{pythia} for subsequent
hadronization. 
Generated events were processed through 
the {\geant}-based~\cite{geant} \dzero 
~detector simulation and reconstructed with the same program used for 
collider data. Additional smearing was applied to the   
reconstructed objects to improve the agreement between data and simulation.  
Remaining discrepancies in the description of the object reconstruction and 
identification between the simulation and the data
were taken into account with correction factors derived
by comparing the efficiencies measured in {\mbox{$ Z\rightarrow {\ell^+\ell^-}$}}\
data events to the ones obtained from the simulation. Lepton and jet trigger 
efficiencies derived from data were also applied to the simulated events.   
The fully corrected efficiencies to select 
$\ttbar$ events were found to be (11.6$\pm$1.7)\% and (11.7$\pm$1.9)\% in the 
$\ejets$ and $\mujets$ channel, respectively.
These efficiencies are calculated with respect to all $\ttbar$ final states 
that contain an electron or a muon originating either directly from a $W$ boson or 
indirectly from $W\rightarrow\tau \nu$ decay. The branching fractions of such  
final states are 17.106\% and 17.036\% \cite{PDG2002} for the $\ejets$ and $\mujets$ channels, 
respectively.

The background within the selected samples is dominated by $W$+jets events, which have 
the same signature as $\ttbar$ signal events. The samples also include 
contributions from multijet events in which
a jet is misidentified as an electron ($\ejets$ channel) or in which  
a muon originating from the semileptonic decay of a heavy quark 
appears isolated ($\mujets$ channel). In addition, 
significant $\met$ can arise from fluctuations and mismeasurements of the
jet energies. We call these instrumental backgrounds ``multijet background'' and
we estimated their contribution directly from data, following the ``matrix'' 
method described in Ref.~\cite{Run1WZPRD} with the 
loose and tight samples described above.
The loose sample consists of $N_{s}$ signal events and $N_{b}$ 
multijet background events, where
$N_{s}$ is a combination of $W$+jets and $\ttbar$ events. 
The tight sample consists of  $\varepsilon_{s} N_{s}$ signal events and
$\varepsilon_{b} N_{b}$ multijet background events, where
$\varepsilon_{s}$ and $\varepsilon_{b}$ are the lepton selection  
efficiencies for the tight sample relative to the loose sample, for signal 
and background, respectively. 
We measured $\varepsilon_{s}$ from a combination of $\ttbar$ and $W$+4~jets 
simulated events, and
applied a correction factor derived from 
the comparison of the corresponding efficiency in   
the {\mbox{$ Z\rightarrow {\ell^+\ell^-}$}}\ data and simulation. We obtained 
$\varepsilon_{b}$ from events with {\mbox{$\not\!\!E_T$}} $<10\;\rm GeV$, which
are dominated by multijet background; $\varepsilon_{b}$ was found to be
independent of jet
multiplicity. For the $\ejets$ channel, 
$\varepsilon_{s}=0.82 \pm 0.02$, and $\varepsilon_{b}=0.16 \pm 0.04$. 
For the $\mujets$ channel, $\varepsilon_{s}=0.81 \pm 0.02$, and $\varepsilon_{b}=0.09 \pm 0.03$. 
\par To extract the fraction of \ttbar ~events in the sample we constructed a 
discriminant function that makes use of 
the differences between the kinematic properties of the \ttbar ~events
and the $W+$jets background. We did not need to consider the multijet 
background separately from the $W+$jets background because  
the kinematic properties of these two event types are similar.
We selected the set of variables that provide the best 
separation between signal 
and background, but have the least sensitivity to the dominant systematic  
uncertainties coming from the jet energy calibration and the $W$+jets 
background model.
To reduce the dependence on modeling of soft radiation and underlying event, 
only the four highest $p_T$ jets were used to determine these variables.
The optimal discriminant function was found to be
built from six variables:
{\it i}) $H_{T}$, the scalar sum of the $p_T$ of the four leading jets; 
{\it ii}) $\Delta\phi(\ell,\not\!\!E_T)$, the azimuthal opening 
angle between the lepton and the missing transverse energy;
{\it iii}) $K_{T{\rm min}} = \Delta
R_{jj}^{\rm min}p_T^{\rm min}/E_T^W$, where $\Delta
R_{jj}^{\rm min}$ is the minimum separation in $\eta-\phi$
space between pairs of jets, $p_T^{\rm min}$ is the $p_T$ of the
lower-$p_T$ jet of that pair, and $E_T^W$ is a scalar sum of the lepton
transverse momentum and~${\mbox{$\not\!\!E_T$}}$; 
{\it iv}) 
the event centrality ${\mathcal C}$, defined as the ratio 
of the scalar sum of the $p_T$ of the jets to the 
scalar sum of the energy of the jets;
{\it v}) the event aplanarity ${\mathcal A}$, constructed from the four-momenta of 
the lepton and the jets; and  
{\it vi}) the event sphericity {$\mathcal S$}, constructed from the four-momenta of 
the jets. 
The last two variables characterize the event shape and are defined, for
example, in Ref.~\cite{tensor}.    
\par The discriminant function was built using the method described in Ref.~\cite{topmass}, and 
has the following general form:
\begin{eqnarray}
\label{eq:discr0}
{\mathcal D} &=& \frac{S(x_1,x_2,...)}{S(x_1,x_2,...) +
B(x_1,x_2,...)}\;,
\end{eqnarray}
where $x_1,x_2,...$ is a set of input variables and 
$S(x_1,x_2,...)$ and $B(x_1,x_2,...)$ 
are the probability density functions for the 
$\ttbar$ signal and background, respectively.
Neglecting the correlations between the input variables, the 
discriminant function can be approximated by the expression:  
\begin{eqnarray}
\label{eq:discr1}
{\mathcal D} &=& \frac{\prod_i s_i(x_i)/b_i(x_i)}{\prod_i s_i(x_i)/b_i(x_i) + 1} \;,
\end{eqnarray}
where $s_i(x_i)$ and $b_i(x_i)$ are the normalized distributions of variable $i$ 
for signal and background, respectively.
As constructed, the discriminant function peaks near zero for the background, and near 
unity for the signal. We modeled it  
using simulated $\ttbar$ and $W$+jets events, and a data sample selected by 
requiring that the leptons fail the tight selection criterion, representative of the 
multijet background. A Poisson maximum-likelihood fit of the modeled discriminant
function distribution to that of the data yielded the top quark cross section 
$\sigma_{t\bar{t}}$ and the numbers of $W$+jets  
and multijet background events in the selected data sample. The multijet
background was constrained within errors to the level determined by the matrix method.         
%--------------------------------
\begin{figure*}[tbh] 
\leftline{\includegraphics[width=0.45\textwidth]{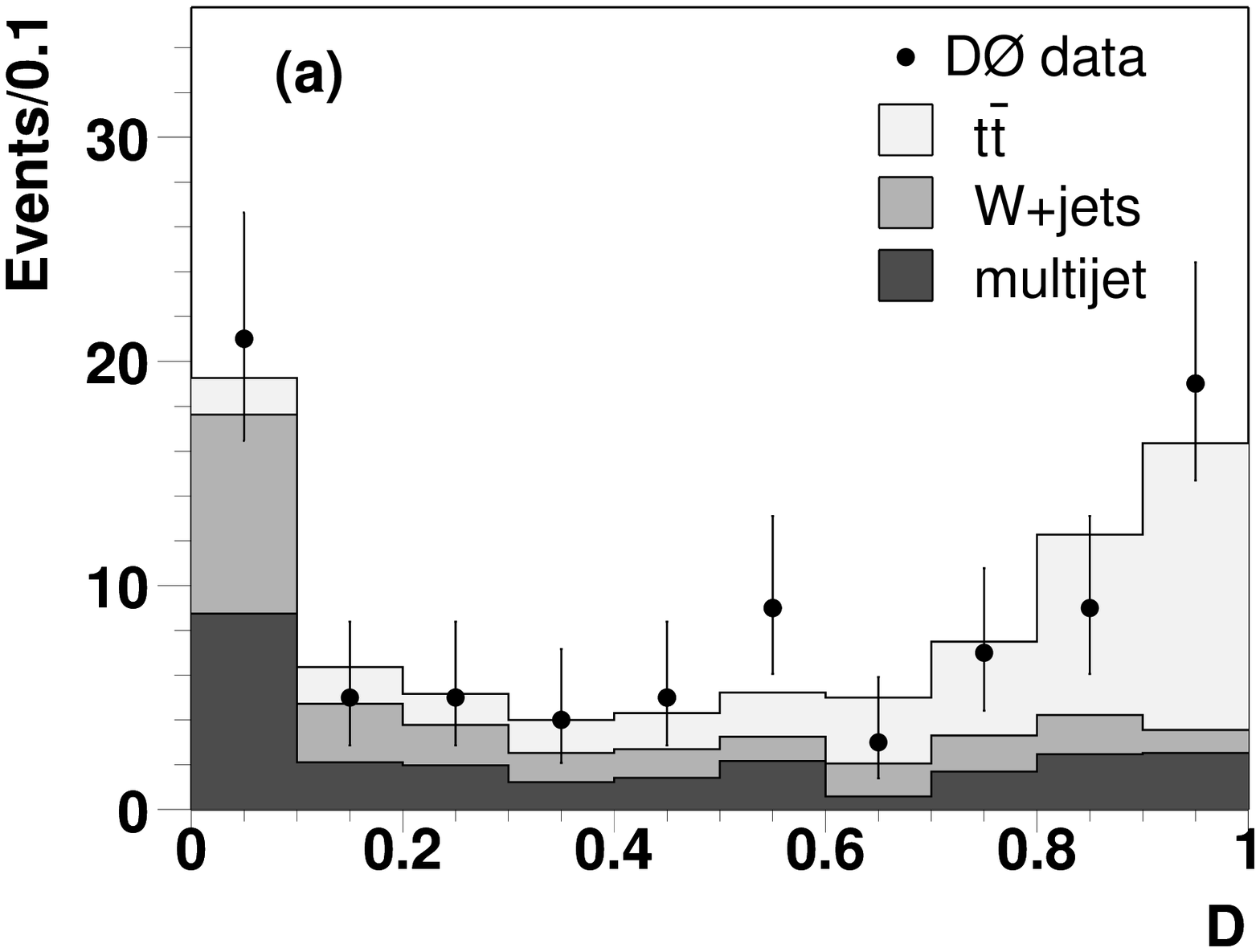}
\includegraphics[width=0.45\textwidth]{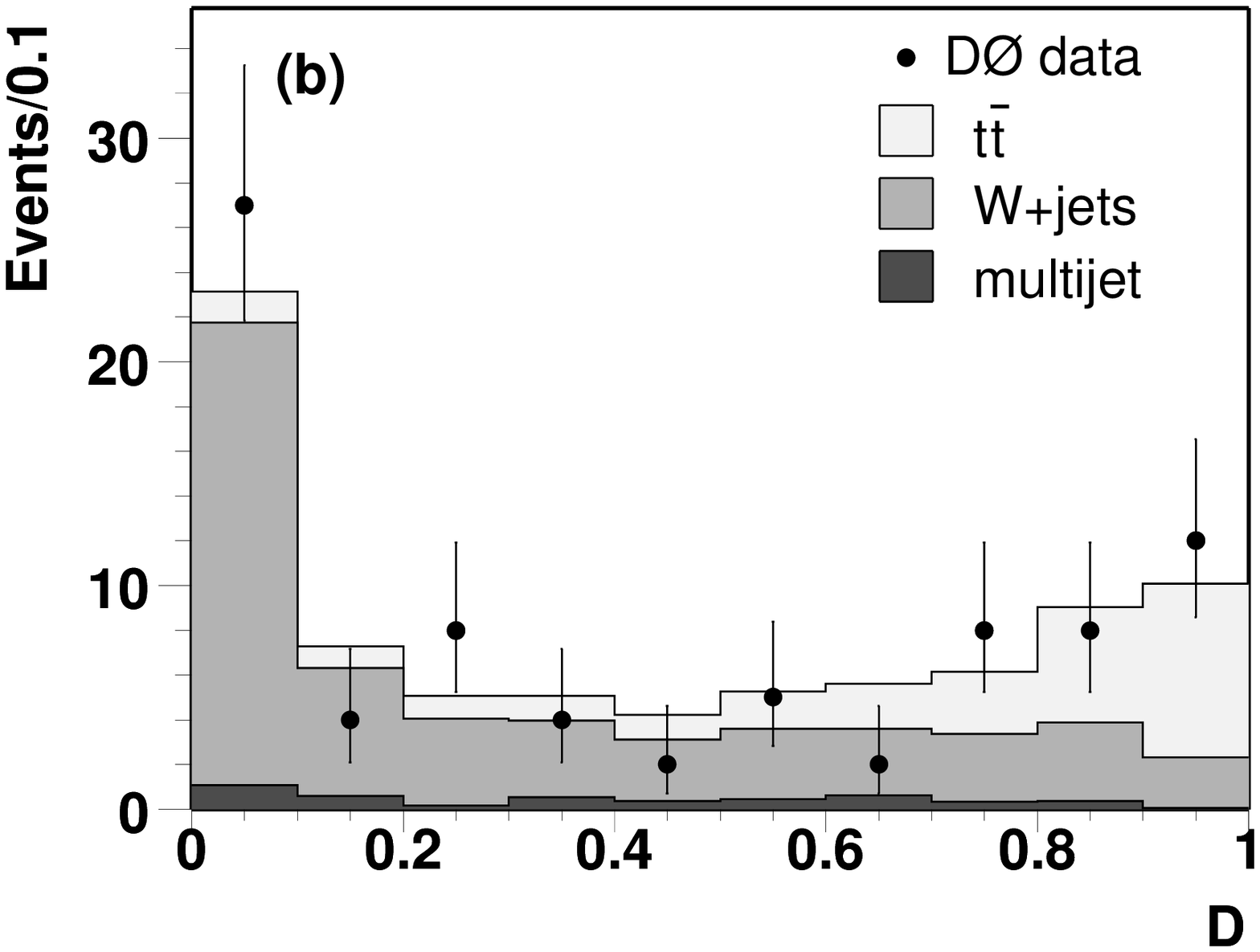}}
\caption{\label{fig:topofit}Discriminant distribution for data 
overlaid with 
the result from a fit of $\ttbar$ signal, and $W$+jets and multijet background 
(a) in the $\ejets$ channel and (b) in the $\mujets$ channel.}
\end{figure*}
%---------------

Figure \ref{fig:topofit} shows the distribution of the discriminant function 
for data along with the fitted contributions from 
$\ttbar$ signal, $W$+jets, and multijet background events. 
The kinematic distributions observed in
lepton+jets events are well described by the sum 
of $\ttbar$ signal, $W$+jets, and multijet background contributions.
An example of this agreement is 
illustrated in Fig.~2 for events selected requiring $\mathcal D < 0.5$,  
dominated by background, and events in the $\ttbar$ signal region of 
$\mathcal D > 0.5$, for
a variable that is not included in the discriminant function, namely, 
the highest jet $p_T$ in the event. 

%--------------------------------
\begin{figure*}[tbh] 
\leftline{\includegraphics[width=0.45\textwidth]{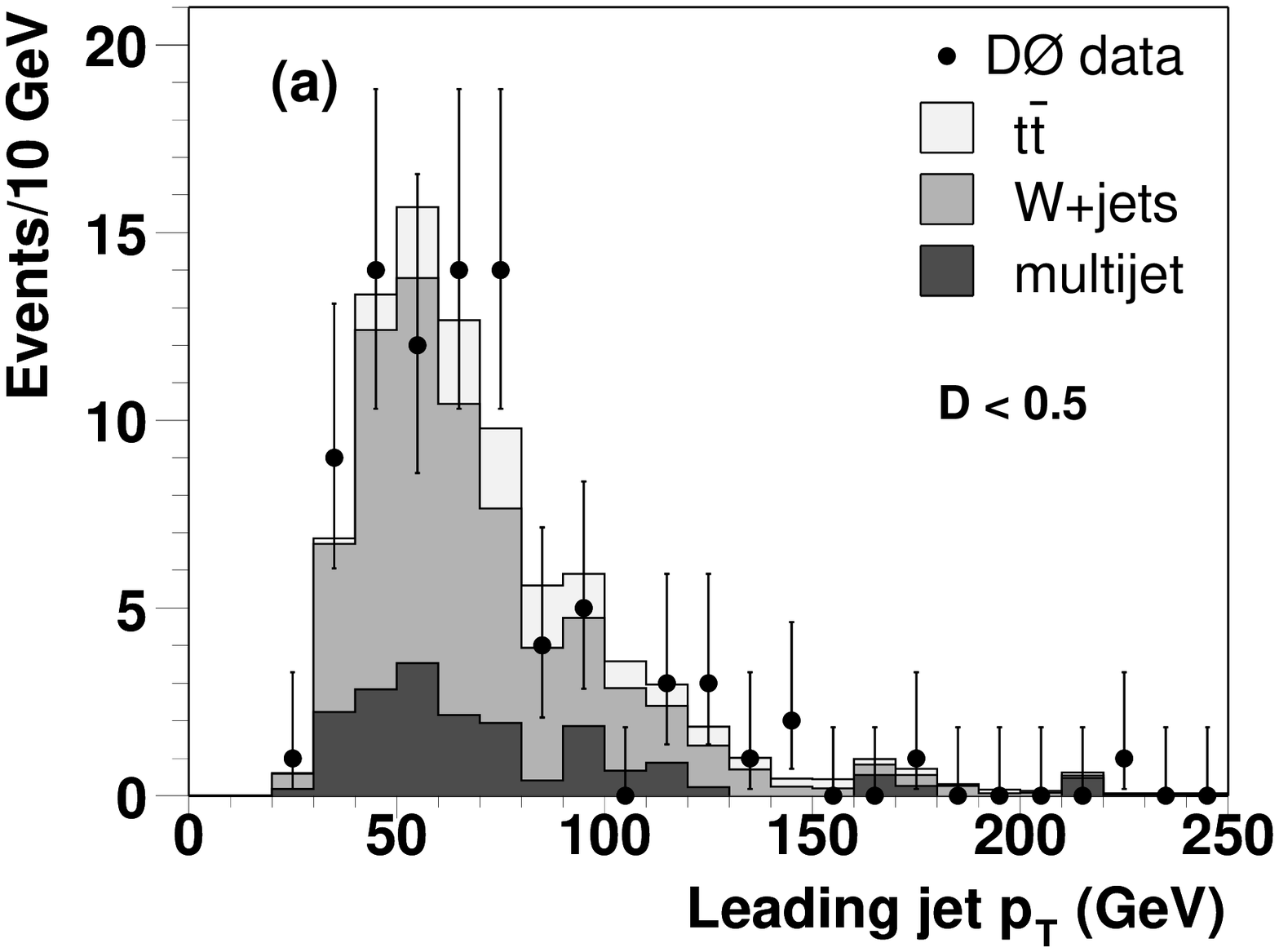}
\includegraphics[width=0.45\textwidth]{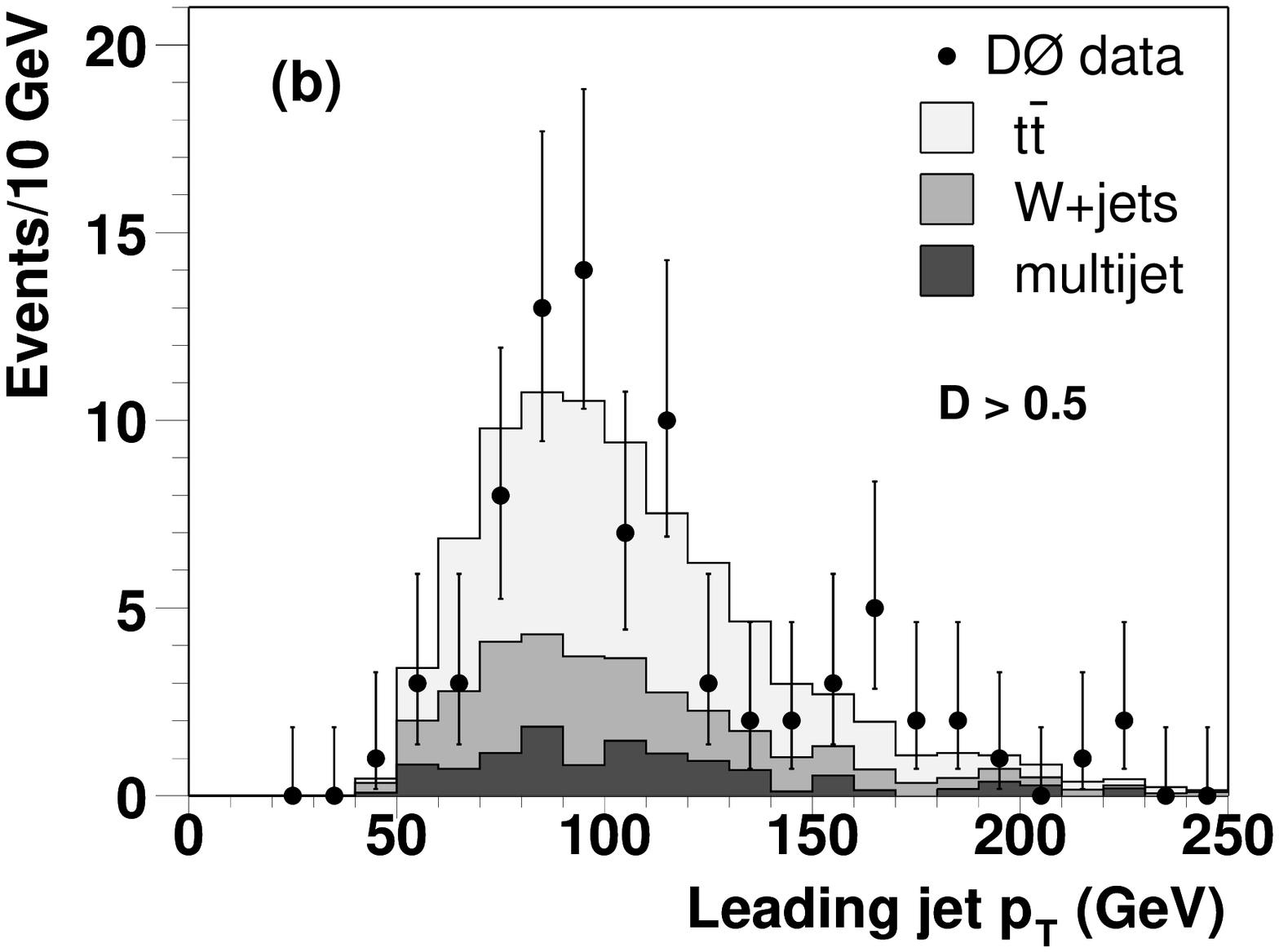}}
\caption{\label{fig:jetpt} Leading jet $p_T$ distribution for $\ljets$ events 
in data with (a) discriminant below 0.5 and (b) discriminant above 0.5, 
overlaid with the result from a fit of $\ttbar$ signal, and $W$+jets and multijet background.}
\end{figure*}
%---------------

The measurement of the $\ttbar$ production cross section
at $\sqrt{s} = $1.96~TeV in each lepton channel separately
yields:
\begin{eqnarray}
{\rm e+jets}&:& \sigma_{t\bar{t}} = 8.2^{+2.1}_{-1.9}\:{\rm (stat)}\:
                                                 ^{+ 1.9}_{- 1.3}\:{\rm (syst)}\:
						 \pm 0.5\:{\rm(lumi)}\:{\rm pb},\nonumber \\
{\rm \mu+jets}&:& \sigma_{t\bar{t}} = 5.4^{+1.8}_{-1.6}\:{\rm (stat)}\:
                                                 ^{+ 1.2}_{- 1.0}\:{\rm (syst)}\:
						 \pm 0.4\:{\rm(lumi)}\:{\rm pb}, \nonumber
\end{eqnarray}
\noindent assuming a top quark mass ($m_t$) of $175\;\rm GeV$. These results 
agree within statistical uncertainties.

The combined cross section was estimated by minimizing the sum of the 
negative log-likelihood functions for each individual channel yielding
\begin{eqnarray}
\sigma_{t\bar{t}} =  6.7^{+1.4}_{-1.3}\:{\rm (stat)}\:
       ^{+1.6}_{-1.1}\:{\rm (syst)}\: \pm0.4\:{\rm(lumi)}\:{\rm pb} \nonumber
\end{eqnarray}
for $m_t=175\;\rm GeV$. 
We treated systematic uncertainties contributing to the error on the event 
selection efficiency and the likelihood fit as 
fully correlated between each other and between the channels. The contributions to the systematic uncertainty from 
the different sources considered in the analysis 
are presented in Table \ref{tab:sys}. The jet energy calibration   
uncertainty dominates, and represents about 
90\% of the total systematic uncertainty on the cross section.      
In addition, a systematic uncertainty of $6.5\%$ from the luminosity
measurement~\cite{d0lumi} has been assigned.
In the top quark mass range of 160 GeV to 190 GeV, the measured
cross section decreases (increases) by $0.11\;\rm pb$ per 1 GeV 
shift of $m_t$ above (below) 175 GeV.

\begin{table}[h]
\caption{\label{tab:sys} Summary of systematic 
uncertainties $\Delta \sigma_{t\bar{t}}$ (pb).}
\begin{ruledtabular}
\begin{tabular}{lccc}
  Source                 & $\ejets$  &   $\mujets$   & $\ell+$jets     \\  
\hline
Lepton identification &  $\pm \;0.3$     & $\pm \;0.2$   &  $\pm \;0.2$	  \\   
Jet energy calibration &  $+\;1.8-1.2$   & $+\;1.0-0.7$  &  $+\;1.4-1.0$ \\   
Jet identification    &  $+\;0.2-0.2$   & $+\;0.2-0.1$  &  $+\;0.2-0.1$ \\   
Trigger               &  $+\;0.1-0.1$   & $+\;0.4-0.3$  &  $+\;0.3-0.2$ \\   
Multijet background   &  $\pm \;0.3$     & $\pm \;0.03$  &  $\pm \;0.2$   \\  
$W$ background model  &  $\pm \;0.2$     & $\pm \;0.4 $  &  $\pm \;0.3$   \\   
MC statistics         &  $\pm \;0.5$     & $\pm \;0.3 $  &  $\pm \;0.3$   \\ 
Other                 &  $\pm \;0.2$     & $\pm \;0.1 $  &  $\pm \;0.2$   \\ \hline
Total                 &  $+\;1.9-1.3$   & $+\;1.2-1.0$  &  $+\;1.6-1.1$ \\   
\end{tabular}
\end{ruledtabular}
\end{table}

In summary, we have measured the top quark pair production cross section in 
{\mbox{$p\bar p$}}\ collisions at {\mbox{$\sqrt{s}$ =\ 1.96\ TeV}}\ in the 
lepton+jets channel. Our measurement is consistent with the SM 
expectation~\cite{SMtheory} which predicts a cross 
section of $6.77 \pm 0.42\;\rm pb$ for a top quark mass of $175\;\rm GeV$.

% acknowledgement_paragraph_r2.tex                4/8/05
%
We thank the staffs at Fermilab and collaborating institutions, 
and acknowledge support from the 
DOE and NSF (USA),
CEA and CNRS/IN2P3 (France),
FASI, Rosatom and RFBR (Russia),
CAPES, CNPq, FAPERJ, FAPESP and FUNDUNESP (Brazil),
DAE and DST (India),
Colciencias (Colombia),
CONACyT (Mexico),
KRF (Korea),
CONICET and UBACyT (Argentina),
FOM (The Netherlands),
PPARC (United Kingdom),
MSMT (Czech Republic),
CRC Program, CFI, NSERC and WestGrid Project (Canada),
BMBF and DFG (Germany),
SFI (Ireland),
A.P.~Sloan Foundation,
Research Corporation,
Texas Advanced Research Program,
Alexander von Humboldt Foundation,
and the Marie Curie Fellowships.
%

%\begin{references}


\begin{thebibliography}{999}
\bibitem{eta} Rapidity $y$ and pseudorapidity $\eta$ are defined as functions 
of the polar angle $\theta$ as 
$y(\theta,\beta) \equiv
{1 \over 2} \ln{[(1+\beta\cos{\theta})/(1-\beta\cos{\theta})]}; 
\eta(\theta) \equiv y(\theta,1)$, where $\beta$ is a ratio of particle momentum
to its energy.
\end{thebibliography}
\end{document}